\documentclass[12pt]{article}
\usepackage{a4wide}
\usepackage{epsfig,graphicx}
\usepackage{amsmath}
\usepackage{amssymb}
\usepackage{array}
\usepackage{slashed}
\newcommand{\beq}{\begin{equation}}
\newcommand{\eeq}{\end{equation}}
\newcommand{\bea}{\begin{eqnarray}}
\newcommand{\eea}{\end{eqnarray}}
\newcommand{\nn}{\nonumber \\}

\newcommand{\OMIT}[1]{{}}

\newcommand\spur{\raise.15ex\hbox{/}\kern-.57em }

\newcommand{\lsim}{
\mathrel{\hbox{\rlap{\hbox{\lower4pt\hbox{$\sim$}}}\hbox{$<$}}}}
\newcommand{\gsim}{
\mathrel{\hbox{\rlap{\hbox{\lower4pt\hbox{$\sim$}}}\hbox{$>$}}}}
\begin{document}

\begin{flushright}
LA-UR-07-7940\\
UWThPh-2008-12\\
July 2008
\end{flushright}
\vspace{2.0 true cm}
\begin{center}
{\Large {\bf 
Electromagnetic effects  in 
$\mbox{\boldmath
$K_{\ell 3}$}$ decays${}^*$   
}}\\
\vspace{2.0 true cm}
{\large Vincenzo Cirigliano${}^a$,  Maurizio Giannotti${}^a$, Helmut Neufeld${}^b$}\\
\vspace{1.0 true cm}
${}^a$   {\sl Theoretical Division, Los Alamos National Laboratory, Los Alamos, NM 87545} \\
\vspace{0.2 true cm}
${}^b$ {\sl Fakult\"at f\"ur Physik, Universit\"at Wien, Boltzmanngasse 5, A-1090 Wien, Austria } \\
%${}^c$ {\sl  } \\
%\vspace{0.2 true cm}
% \today
\end{center}
\vspace{1.5cm}

\begin{abstract}
We  study the radiative corrections to all $K_{\ell 3}$ decay modes 
to leading non-trivial order in the chiral effective field theory, 
working with a fully inclusive prescription on real photon emission.  
We present new results for $K_{\mu 3}$ modes and update  
previous results on $K_{e3}$ modes. 
Our analysis provides 
important theoretical input for the extraction of the CKM element $V_{us}$ 
from $K_{\ell 3}$ decays. 
\end{abstract}

\vfill

$^*$ This work was supported in part by the EU Contract No. 
MRTN-CT-2006-035482, \lq\lq FLAVIAnet". 

\newpage

\section{Introduction}

With the advent of precision measurements  in Kaon physics  
(see ~\cite{flavianet} and references therein), 
$K_{\ell 3}$ decays offer  the opportunity to probe  
charged current weak interactions at unprecedented levels. 
Most notably, with current experimental uncertainties,  $K_{\ell 3}$ decays 
allow one to access the   
Cabibbo-Kobayashi-Maskawa (CKM) quark mixing angle $V_{us}$ at the sub-percent level,  
and also provide competitive probes of lepton universality and the ratios of light quark masses.   
In order to fully exploit  the amazing experimental achievements, 
it becomes mandatory to have theoretical control  
of these decays at the percent level or better.  
This requires quantitative understanding of the vector and scalar $K \to \pi$ form factors 
as well as the electromagnetic (EM)  corrections. 
The framework to analyze the EM corrections is provided by 
Chiral Perturbation Theory (ChPT)~\cite{chpt}, 
the low energy effective field theory (EFT) of QCD, 
extended to include the photon~\cite{photons} and the light leptons ~\cite{Knecht:1999ag}
as active degrees of freedom.
ChPT exploits the  special role of   $\pi, K , \eta$ as Goldstone modes 
associated with the spontaneous  breaking of chiral $SU(3)_L \times SU(3)_R$  symmetry, 
and provides a systematic expansion of the amplitudes  in 
powers of the masses of pseudoscalar mesons and charged leptons 
($p \sim M_{\pi,K,\ell}/\Lambda_\chi$ with $\Lambda_\chi \sim 4\pi F_\pi \sim 1.2$ GeV) 
and the electromagnetic coupling $(e)$. 

In this article  we present results on 
the electromagnetic corrections  to the four $K_{\ell 3}$ decay modes ($K=K^\pm,K^0;   \ell= e,\mu $), 
based on a calculation of the amplitudes to leading non-trivial order in ChPT ($O(e^2 p^2)$). 
For all  modes, we focus on (i)  the  electromagnetic (EM) corrections to the 
Dalitz plot, which are needed 
to extract 
the momentum 
dependence of the $K \to \pi$ vector and scalar  form factors
from the experimental distribution; 
(ii) the integrated radiative correction  to the decay rate, which is a crucial input 
in extracting  the CKM mixing angle $V_{us}$ from $K \to \pi \ell \nu [\gamma]$ decays. 
The theoretical framework for the calculation of electromanetic contributions to $O(e^2 p^2)$
in $K_{\ell 3}$ decays was presented in Ref.~\cite{Cirigliano:2001mk} and 
full numerical results on the $K_{e3}$ modes were given in 
Refs.~\cite{Cirigliano:2004pv}, adopting  
a specific prescription for treating real photon emission 
and a specific factorization scheme for soft photons , 
which results in the {\it partial} inclusion of higher order terms in the chiral expansion. 
The  novel features of the present work can be summarized as follows:

\begin{itemize}

\item  Rather than using the soft-photon factorization procedure of  Ref.~\cite{Cirigliano:2001mk}, 
we work here to {\it fixed chiral order} $e^2 p^2$,  providing the 
complete  corrections to decay 
distributions and total decay rates to $O(e^2 p^0)$. 

\item  We give new results for $K_{\mu 3}$ modes and update our previous 
analysis of  $K_{e3}$ modes. 

\item We use a fully inclusive prescription for real photon emission, which is more
appropriate for comparison with the experimental results.

\item  We update  the  structure-dependent EM correction, 
using the recent estimates   of the relevant low-energy constants (LECs) provided in Refs.~\cite{DescotesGenon:2005pw,mouss-ana}.

\end{itemize}

Preliminary results of our analysis have been made public in conference talks and 
proceedings~\cite{HN-frascati07,VC-kaon07} and should be considered obsolete after the current 
publication.  The paper is organized as follows:  in  
Section~\ref{sect:overview} we give an overview of various contributions to $K_{\ell 3}$ radiative 
corrections and derive the relevant master formula for the corrections to fixed chiral order 
($O(e^2 p^0)$).  In  Section~\ref{sect:results} we present our results for differential and 
total radiative corrections, discussing  their uncertainty.  
In Section~\ref{sect:conclusions} we present our conclusions.

\section{Radiative Corrections to 
$\mbox{\boldmath $K_{\ell 3}$}$ 
decays: overview} 
\label{sect:overview}

\subsection{Generalities on
$\mbox{\boldmath $K_{\ell 3}$}$ 
 decays}

Let us briefly recall  the main features of $K_{\ell 3}$ decays.
The invariant amplitude for the process
$ K (p_K) \to \pi (p_\pi) \, \ell^+ (p_\ell) \, \nu_\ell (p_\nu)$  
reads
\beq
{\cal M} =
\frac{G_{\rm F}}{\sqrt{2}} V_{us}^{*} \
\bar{u} (p_\nu) \, \gamma^\mu \, (1 - \gamma_5) \, v (p_\ell) 
\ C_K 
\ \bigg[ f_{+}^{K \pi} (t) \, (p_K + p_\pi)_{\mu} 
+  f_{-}^{K \pi} (t) \, (p_K - p_\pi)_{\mu} \bigg]  ~,
\label{basic1}
\eeq
where  $C_K  = 1$ for $K^{0}_{\ell 3}$ and  $C_K= 1/\sqrt{2}$ for $ K^{+}_{\ell 3}$ modes. 
The expression in square brackets  corresponds to the matrix element $
\langle \pi (p_\pi) | V_{\mu}^{4-i5} | K (p_K) \rangle $, expressed in
terms of the form factors $f_{\pm}^{K \pi} (t)$, which 
depend on the single variable $t = (p_K - p_\pi)^2$  
and are known to $O(p^6)$ in ChPT~\cite{Post:2001si,Bijnens:2003uy,bijnens07}.
To this order, a number of unknown LECs appear, of which  only a few  can be determined 
experimentally.  
A complete prediction to $O(p^6)$ requires theoretical input beyond ChPT, 
either from analytic approaches~\cite{f0-analytic}  or lattice QCD~\cite{f0-lattice}. 
For  phenomenological applications, it is common to parameterize the form factors 
$f_{+} (t)$ and 
$f_0(t) = f_+ (t) +  t/(M_K^2 - M_\pi^2)  f_{-} (t) $  in terms of slope and 
curvature parameters ~\footnote{See 
Ref~\cite{sternetal} for a dispersive parameterization of the scalar form factor.}   
which can then be measured: 
\bea
  \bar{f}^{K \pi}_{+} (t) \equiv   \frac{f^{K \pi}_{+} (t)}{f_+^{K \pi}(0)}
 &=& 1  + \lambda_{+}  \, \frac{t}{M_{\pi^\pm}^2}   + \frac{1}{2}
\lambda_{+}^{''}  \, \frac{t^2}{M_{\pi^\pm}^4} ~, 
\\
 \bar{f}^{K \pi}_{-} (t) \equiv   \frac{f^{K \pi}_{-} (t)}{f_+^{K \pi}(0)}
  &=&    \frac{M_K^2 - M_\pi^2}{M_{\pi^\pm}^2} \, \left( 
\lambda_0 - \lambda_+  - \frac{\lambda_+^{''}}{2}  \frac{t}{M_{\pi^{\pm}}^2} \right) ~.
\eea

The spin-averaged decay distribution depends on two independent kinematical variables,  
which we choose to be 
\beq z = \frac{ 2 p_\pi \cdot p_K }{M_K^2} = 
\frac{2 E_{\pi}}{M_K}  ~,  \quad y = \frac{ 2 p_K \cdot p_\ell }{M_K^2} =
\frac{2 E_{\ell}}{M_K}  ~, \eeq
where $E_\pi$ ($E_{\ell}$) is the pion (charged lepton) energy in the 
kaon rest frame, and $M_K$ indicates the mass of the decaying kaon. 
Then the distribution (without radiative corrections) reads 
\bea
\frac{d \Gamma^{(0)}}{dy \, d z} &= & 
   \frac{G_{\rm F}^2 \, |V_{us}|^2 \, M_K^5 \, C_K^2 }{128 \, \pi^3} 
\ |f^{K \pi}_{+}(0) |^2  \  \bar{\rho}^{(0)} (y,z) ,
\\
\bar{\rho}^{(0)} (y,z) &= &
 A_1^{(0)} (y,z)  \  |\bar{f}_{+}^{K \pi} (t)|^2  
\, +\, A_2^{(0)}  (y,z)  \  \bar{f}_{+}^{K \pi} (t)   \bar{f}_{-}^{K \pi} (t) 
\, + \,  A_3^{(0)} (y,z)  \  | \bar{f}_{-}^{K \pi} (t)|^2 ~,  
\label{basic2}
\eea
where  the kinematical densities read ($r_\ell = (m_\ell/M_K)^2$  and $r_\pi = (m_\pi / M_K)^2$ ): 
\bea
A_1^{(0)} (y,z) &=& 4 (z + y - 1) (1 - y) 
+ r_\ell (4 y + 3 z - 3) - 4 r_\pi + r_\ell (r_\pi - r_\ell)    ~, \nonumber \\
A_2^{(0)} (y,z) & = & 2 r_\ell (3 - 2 y - z + r_\ell - r_\pi)  ~, \nn
A_3^{(0)} (y,z) & = &  r_\ell ( 1 + r_\pi - z - r_\ell)   ~. 
\eea
In the analysis of $K_{e3}$ decays, the terms proportional to
$A_{2,3}^{(0)}$ can be neglected, being proportional to $r_e \simeq 10^{-6}$. 
Finally, the decay rate reads
\bea
 \Gamma^{(0)}  (K_{\ell 3}) &= &
%{\cal N} 
   \frac{G_{\rm F}^2 \, |V_{us}|^2 \, M_K^5 \, C_K^2 }{128 \, \pi^3}  
\ |f^{K \pi}_{+}(0) |^2 \
I^{(0)}_{K \ell}  (\lambda_i) ~,  
\label{eq:rate0}
\\
I^{(0)}_{K \ell }  (\lambda_i)  &=& 
\int\limits_{{\cal D}_3}  dy \, dz \ \bar{\rho}^{(0)} (y,z)  ~,  
\eea
where the integral extends on the physical domain ${\cal D}_3$
defining the three-body Dalitz plot (see Ref.~\cite{Cirigliano:2001mk} for the explicit definition).

\subsection{Radiative corrections: soft factorization vs  fixed chiral order} 
\label{sect:rad-discussion} 

The above description of differential distributions and decay rates is  
modified by EM effects, which involve the emission of both virtual   
and real photons. 
Short distance electroweak corrections  can be lumped in an overall factor 
$S_{\rm ew} = 1 + \frac{2 \alpha}{\pi} \left( 1 -\frac{\alpha_s}{4 \pi} \right)\times 
\log \frac{M_Z}{M_\rho}  + O (\frac{\alpha \alpha_s}{\pi^2})$, 
which is common to  all  semileptonic
charged-current processes~\cite{Sirlin-sew}. 
Long distance EM corrections to $K_{\ell3}$ decays   
can be studied within ChPT. 
The leading non-trivial corrections to the amplitudes appear to 
$O(e^2 p^2)$  and imply  corrections to 
the form factors, differential distributions, and decay rates starting to 
$O(e^2 p^0)$ ~\footnote{Since  $\bar{u} (p_\nu) \, \gamma^\mu \, (1 - \gamma_5) \, v (p_\ell) \cdot   
 (p_K \pm p_\pi)_\mu  \sim  O(p^2)$, 
EM corrections to $f_{\pm}^{K\pi}$ start at $O(e^2 p^0)$. 
Moreover, since $y,z,r_\ell, r_\pi \sim O(1)$ we can book the densities  $A_{1,2,3}^{(0)}$ 
as quantities of $O(1)$. Therefore,  corrections of 
$O(e^2 p^0)$ to  $f_{\pm}^{K \pi}$ induce 
corrections to the decay distributions and rates of  $O(e^2 p^0)$ (see Eqs.~\ref{basic2}
and~\ref{eq:rate0}).}.

In Ref.~\cite{Cirigliano:2001mk}, it was argued that  long distance EM effects 
can be  taken into account by (i) a universal (i.e. non structure-dependent) 
shift in the densities $A_i^{(0)} (y,z)$  
accompanied by (ii) structure-dependent corrections to the    
form factors $\bar{f}^{K\pi}_{\pm}(t)$.   
This result was obtained by factorizing out of the amplitude  
the universal soft photon corrections~\cite{IR-YFS-W} that are sensitive only to charges, masses, 
and momenta of the particles involved in the decay. 
While this recipe has the benefit of being simple and elegant, 
it inherently mixes different orders in the chiral power counting 
(e.g. the soft-photon  corrections proportional to $f^{K \pi}_{-}(t)$ only appear in 
the EFT  calculation to $O(e^2 p^4)$). 
As a consequence, the resulting decay distribution and rate  
contain not only the full chiral corrections of order $e^2 p^0$ but also 
{\it incomplete} higher order corrections, generated by the factorization procedure. 
Since we are studying a fully photon-inclusive rate,  there are no large logarithms associated 
with the factorized soft-photon corrections and therefore 
the partial higher order corrections 
that are included with the recipe of Ref.~\cite{Cirigliano:2001mk} are {\it not} expected to 
give the dominant contributions to any given order: 
cancellations  with unknown terms  are possible. 
Motivated by this, in the present work we give the 
corrections to decay distributions and 
rates to fixed chiral order, namely  $O(e^2 p^0)$, to which the complete answer is known. 
We shall then use the comparison with the procedure of  Ref.~\cite{Cirigliano:2001mk} 
as a validation of our estimate of the theoretical uncertainty. 

The virtual photon corrections to all  $K_{\ell 3}$ amplitudes ($K=K^{\pm}, K^0$ and $\ell = e,\mu$) 
are known to $O(e^2 p^2)$~\cite{Cirigliano:2001mk, Cirigliano:2004pv}, 
while the real photon emission was worked out explicitly in those references 
only for $K_{e3}$ modes (and only for a specific prescription on the treatment of real  photon 
emission~\cite{ginsberg}). 
Our goal here is to provide  a unified discussion of radiative corrections of $O(e^2 p^0)$ 
to  all $K_{\ell 3}$ decay rates, working with the fully inclusive prescription on real photon emission. 
We  now sketch the derivation of the corrections induced by virtual and real photon 
emission to {\it fixed chiral order} [$O(e^2 p^0)$],  and how they combine into a master formula  for 
the inclusive rate.

\subsubsection{Virtual photons} 

One-loop amplitudes involving virtual photons, together with the associated local counterterm 
contributions,   induce  an effective  correction  of $O(e^2 p^0)$   to  
the QCD form factors $f_{\pm}^{K \pi}$, of the form:
\bea
f_{+}^{K\pi} (t) & \to   & f_{+}^{K\pi} (t) 
+  \delta f_{+}^{K \pi} (v) 
+ \frac{\alpha}{4 \pi} \, \Gamma_c (v, m_\ell^2, M_c^2 ; M_\gamma^2) ~,
\\
f_{-}^{K\pi} (t) & \to   & f_{-}^{K\pi} (t)  +  \delta f_{-}^{K \pi} (v) ~,
\eea
where $M_c$ is the relevant charged meson mass and 
$v = u \equiv  (p_K - p_\ell)^2$  for $K^{\pm}$ decays while 
$v= s \equiv  (p_\pi + p_\ell)^2$ for $K^0$ decays.
The function $ \Gamma_c (v, m_\ell^2, M_c^2 ; M_\gamma^2)$~\cite{Cirigliano:2001mk} 
encodes the universal soft photon virtual corrections and 
is infrared divergent (thus it depends explicitly on the 
infrared regulator $M_\gamma$). 
On the other hand, the corrections  $\delta f_{\pm}^{K \pi} (v)$  encode 
structure dependent effects through one-loop corrections and chiral  low-energy
constants~\cite{Cirigliano:2001mk}.  
We report their expressions in Appendix~\ref{app:loops} in terms of  functions 
defined in Ref.~\cite{Cirigliano:2001mk}.  

Keeping in mind the chiral properties of $f_{\pm}^{K\pi}(t)$, namely that 
$f_+ (t) =  1 + O(p^2) $ and  
$f_- (t) =  O(p^2)$, 
the effect of virtual corrections to leading order in ChPT amounts to the following 
$O(e^2 p^0)$ shift to the differential distribution of Eq.~\ref{basic2}:   
\bea
\delta \bar{\rho}^{\rm EM-virtual} (y,z)  &=&   
 A_1^{(0)} (y,z)   \cdot  \left[ 
 2 \ \delta f_{+}^{K\pi} (v)  + \frac{\alpha}{2 \pi}  \Gamma_c (v, m_\ell^2, M_c^2 ; M_\gamma^2) 
 \right]  
\nonumber \\
 & + & 
A_2^{(0)} (y,z)  \cdot  \delta f_{-}^{K\pi} (v) \ ~.
%  \Big]
\label{eq:drhov}
\eea

\subsubsection{Real photons}

It is  well known  that  only the inclusive sum of $K \to \pi \ell \nu$ and
$K \to \pi \ell \nu + n\,  \gamma$ (with $n=1,2,...$) decay rates is infrared (IR) finite and  observable.  
In the chiral power counting, 
the leading contribution to the radiative amplitudes 
$K (p_K) \to \pi (p_\pi) \, \ell^+ (p_\ell) \, \nu_\ell (p_\nu) \gamma (k)$  
is of  $O(e p)$: this  is  all we need 
for the analysis of the  rates to $O(e^2 p^0)$. 
To this order the  radiative amplitudes read: 
\bea
{\cal M}_\gamma (K^+ \to \pi^0 \ell^+ \nu_\ell \gamma) &=&
\frac{e \, G_{\rm F}}{\sqrt{2}} V_{us}^{*}  \, C_{K^+}  \times 
\nn
&  & \!\!\! \!\!\! \!\!\! \!\!\! \!\!\! \!\!\! \!\!\! \!\!\! \!\!\!\!\!\!
\bar{u}(p_\nu) 
 \bigg[(1+\gamma_5)  \  (2 \, \slashed{p}_\pi   - m_\ell)   
\left(\frac{\epsilon^* \cdot p_\ell}{k \cdot p_\ell} -  \frac{\epsilon^* \cdot p_K}{k \cdot p_K} + 
\frac{\slashed{k} \slashed{\epsilon}^* }{2 k \cdot p_\ell}
 \right) 
\bigg]  v(p_\ell)~,
\label{eq:Mgammap}
\\
{\cal M}_\gamma (K^0 \to \pi^- \ell^+ \nu_\ell \gamma) &=&
\frac{e \, G_{\rm F}}{\sqrt{2}} V_{us}^{*} \,  C_{K^0}  \times  
\nn
&  & \!\!\! \!\!\! \!\!\! \!\!\! \!\!\! \!\!\! \!\!\! \!\!\! \!\!\!\!\!\!
\bar{u}(p_\nu) 
 \bigg[(1+\gamma_5)  \  (2 \, \slashed{p}_K   +  m_\ell)   
\left(\frac{\epsilon^* \cdot p_\ell}{k \cdot p_\ell} -  \frac{\epsilon^* \cdot p_\pi}{k \cdot p_\pi} + 
\frac{\slashed{k} \slashed{\epsilon}^* }{2 k \cdot p_\ell}
 \right) 
\bigg]  v(p_\ell)~.
\label{eq:Mgammaz}
\eea
The resulting correction to the  differential (or total) decay rate  can be calculated from 
\bea
d \Gamma (K \to \pi \ell \nu \gamma) &=&   \frac{1}{2 M_K}  
\sum_{\rm pol}  \big| {\cal M}_\gamma \big|^2  \  \frac{d \Omega_{\pi \ell \nu \gamma}}{(2 \pi)^8} ~, 
\label{eq:dgammarad}
\\
d \Omega_{\pi \ell \nu \gamma}  &=& \prod_{i=\pi, \ell, \nu, \gamma} \ 
\frac{d^3 p_i}{2 \, p_i^0}  \ \delta^{(4)}  (p_K - p_\pi -  p_\ell - p_\nu - k)  ~,
\eea
where $d \Omega_{\pi \ell \nu \gamma}$ is the 4-body Lorentz invariant phase space.

The square modulus of the radiative amplitude can be decomposed into the sum of an IR singular term 
($T_{\rm IR}$)
and an IR finite inner bremsstrahlung term ($T_{\rm IB}$), as follows
\bea
\sum_{\rm pol}  \ |{\cal M}_\gamma  (K^0 \to \pi^- \ell^+ \nu_\ell \gamma) | ^2 &=&  
\frac{e^2 G_F^2 |V_{us}|^2  C_{K^0}^2}{2}   \  \Big[  T^{K^0 \ell}_{\rm IR} \ +\ T^{K^0 \ell}_{\rm IB} \Big] ~,
\\
\sum_{\rm pol} \ |{\cal M}_\gamma  (K^+ \to \pi^0 \ell^+ \nu_\ell \gamma) | ^2 &=&  
\frac{e^2 G_F^2 |V_{us}|^2  C_{K^+}^2}{2}  \ \Big[  T^{K^+ \ell}_{\rm IR} \ +\ T^{K^+ \ell}_{\rm IB} \Big] ~,  
\eea
with the IR singular pieces given by:
\bea
T_{IR}^{K^0 \ell} &=&  4  \, M_K^4 \, A_1^{(0)} (y,z) \nonumber \\
&\times&  \Bigg[ 
- \frac{M_\pi^2}{(k \cdot p_\pi + \frac{M_\gamma^2}{2})^2}
- \frac{m_\ell^2}{(k \cdot p_\ell + \frac{M_\gamma^2}{2})^2}
+  \frac{2 p_\pi \cdot p_\ell}{(k \cdot p_\pi + \frac{M_\gamma^2}{2}) (k \cdot p_\ell + \frac{M_\gamma^2}{2})}
\Bigg] ~,
\\
& & \nonumber \\
T_{IR}^{K^+ \ell} &=&  4  \, M_K^4 \, A_1^{(0)} (y,z) \nonumber \\
&\times&  \Bigg[ 
- \frac{M_K^2}{(k \cdot p_K - \frac{M_\gamma^2}{2})^2}
- \frac{m_\ell^2}{(k \cdot p_\ell + \frac{M_\gamma^2}{2})^2}
+  \frac{2 p_K \cdot p_\ell}{(k \cdot p_K - \frac{M_\gamma^2}{2}) (k \cdot p_\ell + \frac{M_\gamma^2}{2})}
\Bigg] ~.
\eea
The terms  $T^{K \ell}_{\rm IR}$   generate an infrared divergence when 
integrated over the soft photon region of the four-body phase space, 
while the remaining terms are IR finite.  
Integrating over all variables except $y$ and $z$,  
leads to a correction to the Dalitz plot density of the form  
\beq
\delta \bar{\rho}^{\rm EM-real} (y,z) =  
A_1^{(0)} (y,z)   \cdot   \frac{\alpha}{\pi} \, I_0 (y,z; M_\gamma)
 + \Delta_1^{\rm IB} (y,z)
 ~, 
\eeq
where   $I_0 (y,z; M_\gamma)$   arises from $T_{\rm IR}$ 
while  $\Delta^{\rm IB}_{1}$ from $T_{\rm IB}$. 
The functions   $I_0 (y,z; M_\gamma)$  and 
$\Delta^{\rm IB}_1 (y,z)$  depend   on the cut on  
hard real photon emission. 
Results based on integrating over all kinematically allowed photon energies 
are known analytically~\cite{Cirigliano:2001mk, Cirigliano:2004pv,ginsberg}. 

Finally, although both $I_0 (y,z; M_\gamma)$  and
$ \Gamma_c (v, m_\ell^2, M_c^2 ; M_\gamma^2)$ 
are individually infrared divergent, they 
combine into the IR finite function 
\beq
\Delta^{\rm IR}  (y,z)  =  \frac{\alpha}{\pi} \  \left[ I_0 (y,z; M_\gamma) +
\frac{1}{2} \Gamma_c  (v,m_\ell^2, M^2; M_\gamma^2) \right]~, 
\eeq
and we end up with a finite correction to the Dalitz plot density:
\begin{equation}
\delta \bar{\rho}^{\rm EM} (y,z) =
A_1^{(0)} (y,z)   \cdot   
\Big[  \Delta^{\rm IR} (y,z) +   2 \ \delta f_{+}^{K\pi} (v)   \Big]  \ + \ 
\Delta^{\rm IB}_1 (y,z)    
+ A_2^{(0)} (y,z)  \cdot  \delta f_{-}^{K\pi} (v) ~.
\label{eq:drhotot}
\end{equation}

\subsection{Master Formula}

After inclusion of both long-distance and short distance ($S_{\rm ew}$) radiative 
corrections, the  differential decay distribution reads:
\beq
\frac{d \Gamma}{dy \, d z} =  
   \frac{G_{\rm F}^2 \, |V_{us}|^2 \, M_K^5 \, C_K^2 }{128 \, \pi^3} 
 \ S_{\rm ew}  \  |f^{K \pi}_{+}(0) |^2 
\ \Big[  \bar{\rho}^{(0)} (y,z)  \ + \ \delta \bar{\rho}^{\rm EM} (y,z)  
\Big]~. 
\eeq
This expression should be the basis to properly determine experimentally 
the momentum-dependence of the QCD form factors $\bar{f}_{\pm}(t)$ 
appearing in $\bar{\rho}^{(0)} (y,z)$. 

Corrections to the decay rate are obtained by integrating over the variables 
$y$ and $z$.  In the case of fully inclusive prescription on the 
radiated photon, the real photon EM correction should be integrated 
not only over the 3-body region ${\cal D}_3$ 
but on the whole region ${\cal D}_4$  allowed  by 4-body kinematics. 
Taking into account all these corrections,
 the master formula for $K_{\ell 3}$ decay rates reads: 
\beq
\Gamma (K_{\ell 3[\gamma]})  = 
 \frac{G_{\rm F}^2 \, |V_{us}|^2 \, M_K^5 \, C_K^2 }{128 \, \pi^3} 
\ S_{\rm ew}  \  |f^{K^0  \pi^-}_{+}(0) |^2  \ I^{(0)}_{K \ell} (\lambda_i) \ 
 \Big[ 1 + \delta_{\rm EM}^{K\ell} + \delta_{SU(2)}^{K \pi}   \Big]~,  
\label{eq:master}
\eeq
where the strong isospin breaking correction is 
\beq
\delta_{SU(2)}^{K \pi} \equiv  \left( \frac{f_{+}^{K \pi}(0)}{f_{+}^{K^0 \pi^-}(0)} \right)^2  \ -\ 1
\eeq
and the EM radiative correction 
\beq
 \delta_{\rm EM}^{K \ell}  =  \delta_{\rm EM}^{K \ell} ({\cal D}_3) +
\delta_{\rm EM}^{K \ell} ({\cal D}_{4-3})
\eeq
receives contributions from the 3-body and 4-body kinematical regions:
\bea
 \delta_{\rm EM}^{K \ell} ({\cal D}_3) 
 &=& \frac{1}{I^{(0)}_{K \ell} (\lambda_i)} \cdot  
 \int_{{\cal D}_3} \, dy \, dz \  \delta  \bar{\rho}^{\rm EM} (y,z) ~, 
\label{eq:deltaI3} 
\\
 \delta_{\rm EM}^{K \ell} ({\cal D}_{4-3}) 
 &=&    \frac{1}{I^{(0)}_{K \ell} (\lambda_i)} \cdot   
  \frac{\alpha}{2 \,  \pi^4 M_K^6} \,   \int_{{\cal D}_{4-3}}  
\ d \Omega_{\pi \ell \nu \gamma} \   \left(T_{\rm IR}^{K\ell} + T_{\rm IB}^{K\ell} \right) ~. 
\label{eq:deltaI4}  
\eea
Note that Ginsberg's prescription~\cite{ginsberg} for real photon emission, which was adopted in 
Refs~\cite{Cirigliano:2001mk,Cirigliano:2004pv}, amounts to discarding 
the integral in Eq.~\ref{eq:deltaI4}.

\section{Results and discussion}
\label{sect:results}

%%%%%%%%%%%%%%%%%%%%%%%%%%%%%%%%
\begin{figure}[!t]
\centering
\begin{picture}(300,120)  
%%%%
\put(240,50){\makebox(50,50){\epsfig{figure=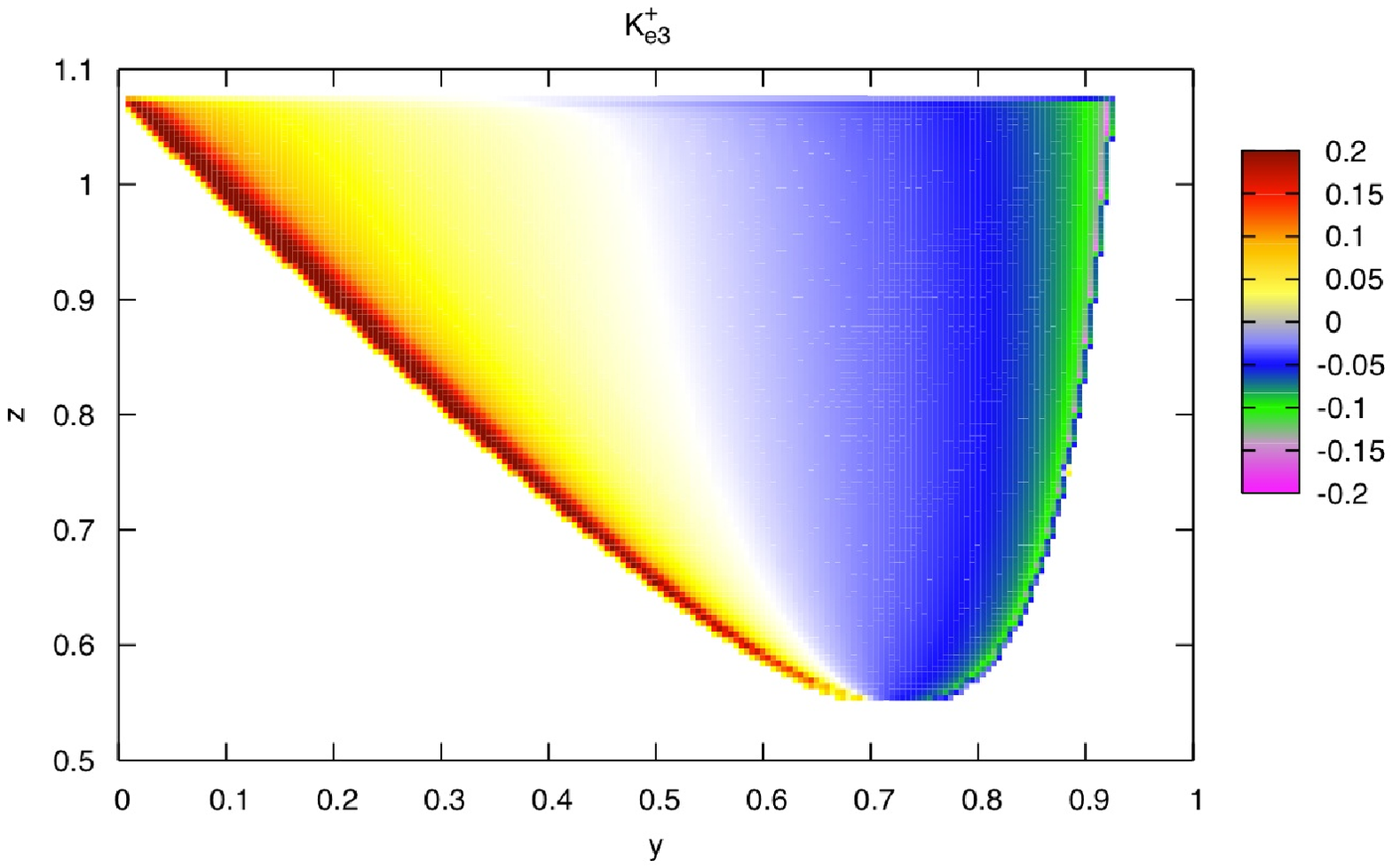,width=10cm}}}
\put(0,50){\makebox(50,50){\epsfig{figure=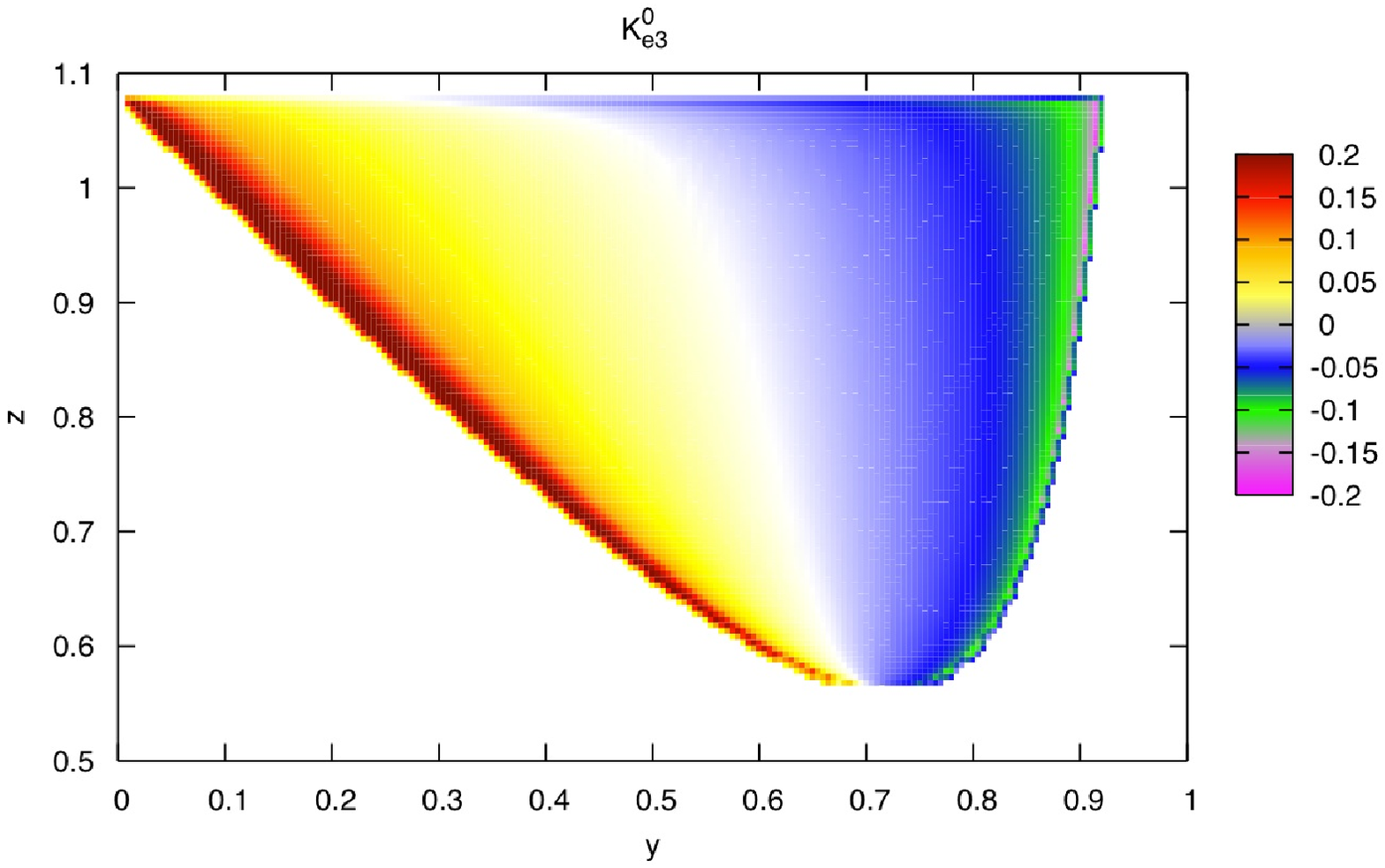,width=10cm}}}
%%%%
%\put(40,0){(a)}
%\put(140,0){(b)}
%\put(240,0){(c)}
%%%%%
\end{picture}
\caption{ 
Density plot of the EM correction to the differential distribution
($\delta \bar{\rho}^{\rm EM}(y,z) /\bar{\rho}^{(0)}(\lambda_i) (y,z)$) 
of  $K^0_{e3}$ (left panel) and $K^{\pm}_{e3}$ (right panel).  
\label{fig:fig1}
}
\end{figure}
%%%%%%%%%%%%%%%%%%%%%%%%%%%%%%%%

The main outcome of our analysis are the differential corrections 
$\delta \bar{\rho}^{\rm EM} (y,z)$ to the 
Dalitz plot density 
 (Eq.~\ref{eq:drhotot})
and the integrated corrections  
$\delta_{\rm EM}^{K \ell} ({\cal D}_3)$ 
and 
$\delta_{\rm EM}^{K \ell} ({\cal D}_{4-3})$ 
(Eqs.~\ref{eq:deltaI3} and \ref{eq:deltaI4} respectively). 
$\delta \bar{\rho}^{\rm EM} (y,z)$  is known analytically through the work of 
Refs.~\cite{Cirigliano:2001mk, Cirigliano:2004pv,ginsberg}. 
The integration needed to calculate 
$\delta_{\rm EM}^{K \ell} ({\cal D}_3)$  
 has been performed with the Gauss quadrature method.
On the other hand, 
the integration needed to calculate 
$\delta_{\rm EM}^{K \ell} ({\cal D}_{4-3})$ 
has been performed with a Monte Carlo technique 
based on the RAMBOS event generator~\cite{rambos}. 

In order to give numerical results, we have to specify the input parameters. 
The differential  and integrated EM corrections to $O(e^2 p^0)$ depend on a number 
of LECs of ChPT.  
The EM LECs are given by convolutions of appropriate QCD correlators  
with known kernels.   They have been recently estimated in 
Refs.~\cite{DescotesGenon:2005pw,mouss-ana} 
by replacing the QCD correlators with meromorphic approximants, in the spirit of 
the large-$N_C$ expansion. 
We use the results of  Refs.~\cite{DescotesGenon:2005pw,mouss-ana} for our central 
values, and conservatively assign $100 \%$ fractional uncertainty to the LECs.

The integrated corrections 
$\delta_{\rm EM}^{K \ell} ({\cal D}_3)$ 
and 
$\delta_{\rm EM}^{K \ell} ({\cal D}_{4-3})$ also depend, through the normalization 
factor $I_{K \ell}^{(0)} (\lambda_i)$, 
on the slope and curvature of the scalar and 
vector form factors, namely $\lambda_+$,  $\lambda_{+}^{''}$, and $\lambda_0$. 
For these quantities we  use the results of a global fit to all consistent 
experimental data (i.e. without including the NA48 result)
as reported in  Ref.~\cite{flavianet}:
$\lambda_+ =  (25.0   \pm  0.8) \cdot 10^{-3}$, 
$\lambda_+^{''} =  (1.6   \pm  0.4) \cdot 10^{-3}$, 
$\lambda_0 =  (16.0   \pm  0.8) \cdot 10^{-3}$. 
The choice of this reference set of input parameters 
is very simple but  somewhat inconsistent 
as  the charged and neutral K parameters are distinguished by 
strong isospin breaking and EM effects~\cite{bijnens07,helmut}. 
However,  one should keep in mind that the 
uncertainty on $I_{K \ell}^{(0)} (\lambda_i)$ induced by the $\lambda_i$ 
is below the percent   level and is completely negligible in the analysis of  $\delta_{\rm EM}^{K \ell}$
(although it has some impact on the extraction of $V_{us}$ through Eq.~\ref{eq:master}).
In Table~\ref{tab:chiral} we provide the
$I_{K \ell}^{(0)} (\lambda_i)$  corresponding to our choice $\lambda_i$, 
so the reader can easily convert our results for  $\delta^{K \ell}_{\rm EM}$  
to any choice of slope parameters  with a simple re-scaling. \\

%%%%%%%%%%%%%%%%%%%%%%%%%%%%%%%%
\begin{figure}[!t]
\centering
\begin{picture}(300,120)  
%%%%%%%%
\put(240,50){\makebox(50,50){\epsfig{figure=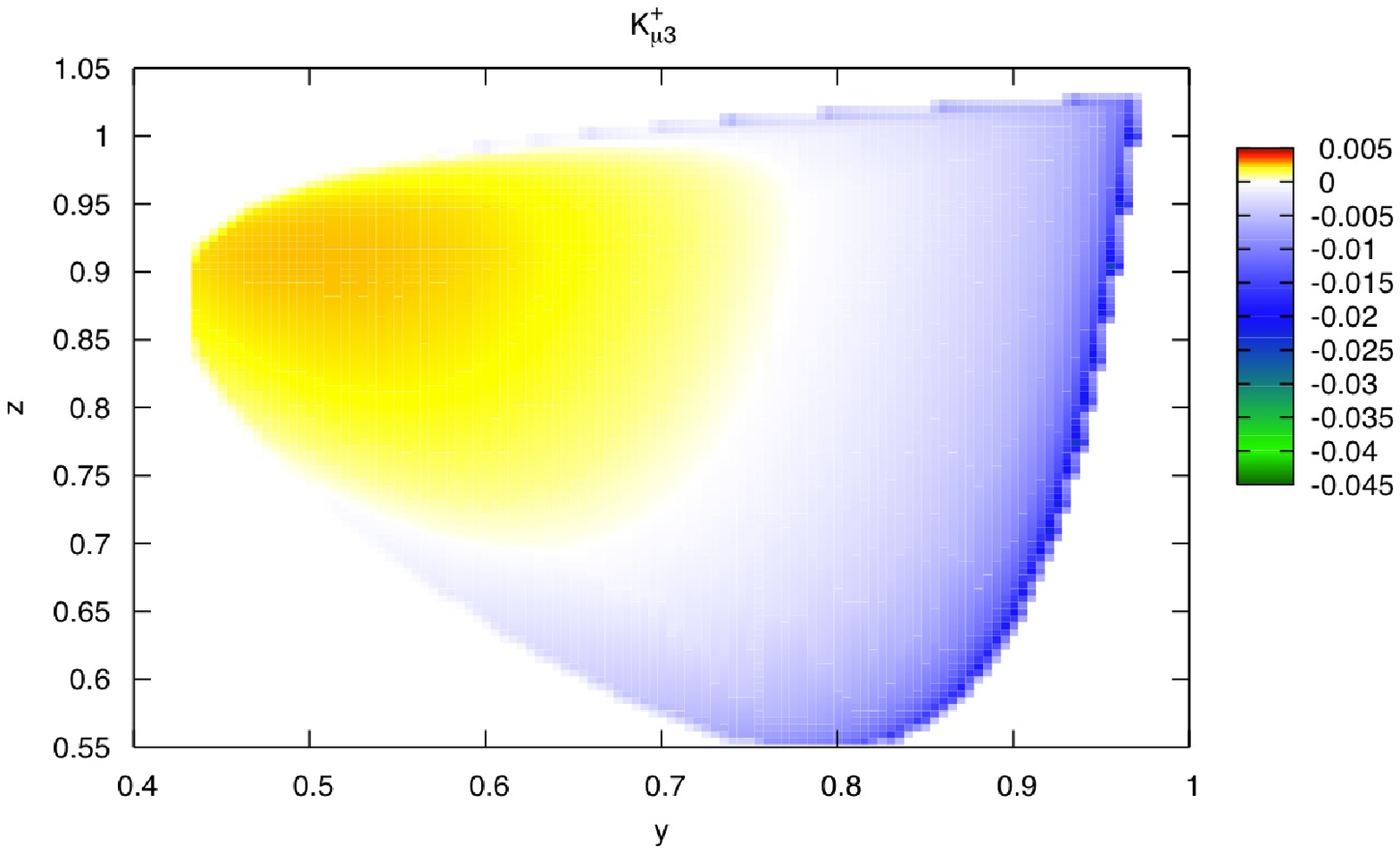,width=10cm}}}
\put(0,50){\makebox(50,50){\epsfig{figure=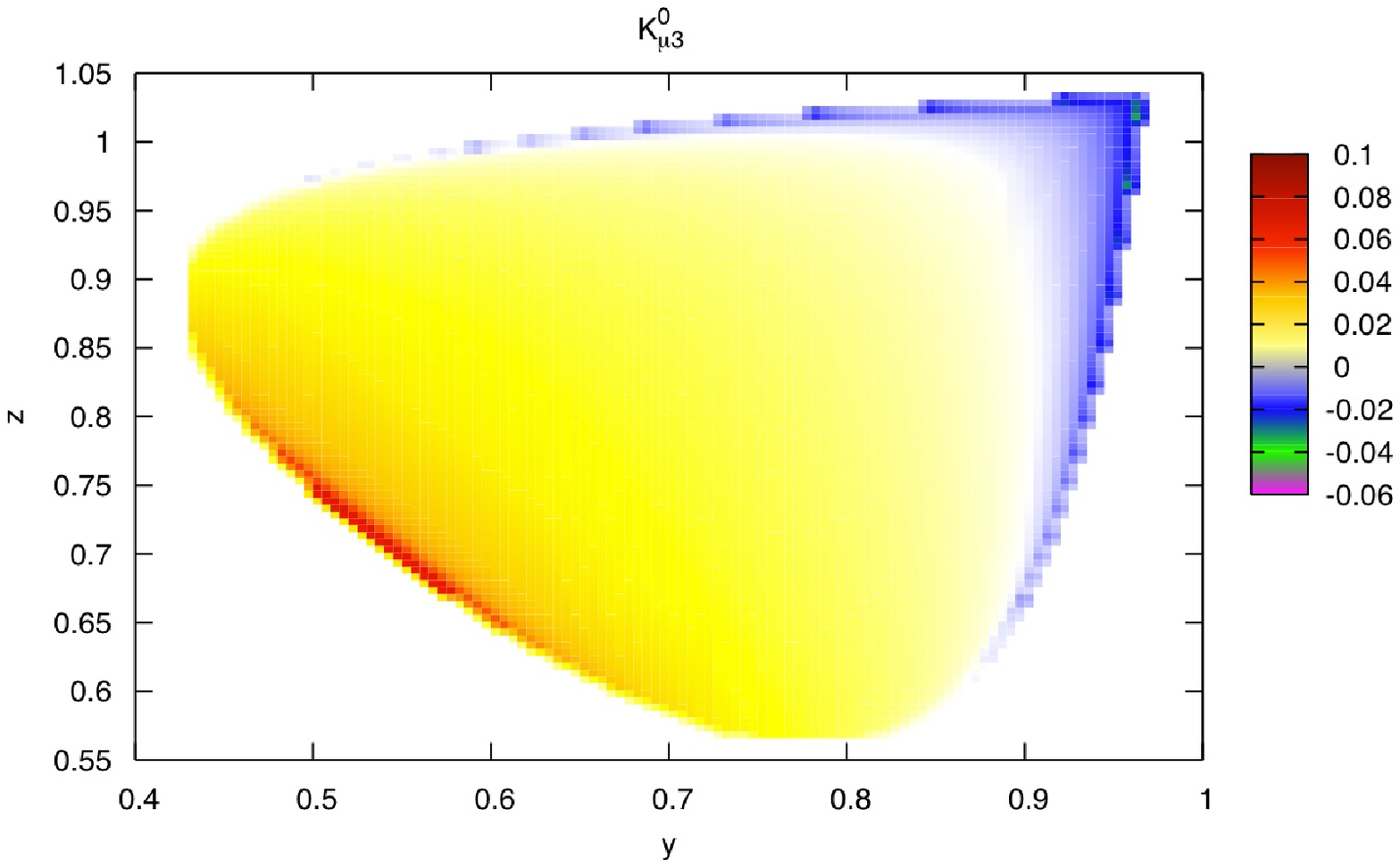,width=10cm}}}
%%%%%%%%
%\put(195,32){{$\frac{1}{2}  z_{\ell}^{e^2} \times $}}
%%%%
%\put(40,0){(a)}
%\put(140,0){(b)}
%\put(240,0){(c)}
%%%%%
\end{picture}
\caption{ 
Density plot of the EM correction to the differential distribution
($\delta \bar{\rho}^{\rm EM}(y,z) /\bar{\rho}^{(0)}(\lambda_i) (y,z)$) 
of  $K^0_{\mu3}$ (left panel) and $K^{\pm}_{\mu3}$ (right panel).  
\label{fig:fig2}
}
\end{figure}
%%%%%%%%%%%%%%%%%%%%%%%%%%%%%%%%

\subsection{Corrections to the Dalitz plot} 

In Figures~\ref{fig:fig1} and \ref{fig:fig2} we show a density plot of 
the ratio  $\delta \bar{\rho}^{\rm EM} /\bar{\rho}^{(0)}(\lambda_i)$  
for  $K^{0}_{e3}$, $K^{\pm}_{e3}$,  $K^{0}_{\mu3}$ and 
$K^{\pm}_{\mu 3}$  as a function of the variables $y,z$, corresponding to the 
input on $\lambda_{i}$ and LECs  specified in the 
previous subsection ~\footnote{Numerical tables for these corrections are 
available from the authors upon request. }. 
The theoretical uncertainty on the LECs and higher order corrections induces 
an overall  uncertainty of  about $\pm 0.3\%$  in 
$\delta \bar{\rho}^{\rm EM}/\bar{\rho}^{(0)}(\lambda_i)$. 
It is important to notice that  the correction to the Dalitz distribution can be locally  
large ($O(10 \%)$) and  does not have definite sign, 
implying possible cancellations in the integrated total EM correction.

\subsection{Corrections to the decay rates} 

Table~\ref{tab:chiral} summarizes the numerical results 
of the long-distance radiative corrections to fixed order  $e^2 p^0$, 
obtained using the central values for the LECs, slopes, and curvature of the 
form factors  as described above. 

Two prominent features of the  results in Table~\ref{tab:chiral} can be  understood on 
a qualitative level.  First,  the EM  corrections to the neutral $K$ decays are 
expected to be positive and sizable on account of the Coulomb final state interaction term 
between  $\ell^+$ and $\pi^-$,  that produces a correction factor 
of  $\pi \alpha/v^{\rm rel}_{\ell^+ \pi^-} \sim 2 \% $ over  most of the Dalitz plot.  
While the exact correction and the relative 
size of $K^0_{\mu3}$ and $K^0_{e3}$ depend on other effects such as the emission 
of real photons, the qualitative expectation based on Coulomb interaction  is confirmed by the detailed calculation.     
Second,  the large hierarchy  
$\delta^{K \mu}_{\rm EM} ({\cal D}_{4-3}) \ll   \delta^{K e}_{\rm EM} ({\cal D}_{4-3})$  
admits a simple interpretation in terms of {\it bremsstrahlung} off the charged lepton 
in the final state.  The probability of emitting soft photons is a function of the lepton velocity $v_{\ell}$
which becomes logarithmically singular as $v_{\ell} \to 1$, thus enhancing the electron emission. 
For typical values of $v_{\ell}$ in ${\cal D}_{4-3}$,  the 
semiclassical emission probability~\cite{jackson} implies 
$\delta^{K e}_{\rm EM} ({\cal D}_{4-3}) / \delta^{K \mu}_{\rm EM} ({\cal D}_{4-3}) \sim 20 \to 40$.  

The theoretical uncertainty to be assigned to
$\delta_{\rm EM}^{K \ell}$  arises from two sources:
the input parameters (LECs, $\lambda_i$) appearing in the 
$O(e^2 p^0)$ correction and unknown higher order terms in the EFT expansion, 
starting at $O(e^2 p^2)$. For the parametric uncertainty we find that:

\begin{itemize}
\item    Experimental errors  on the form factor parameters 
$\lambda_{+}$,  $\lambda_{+}^{''}$, and $\lambda_0$ induce 
a fractional  uncertainty in  $I_{K \ell}^{(0)} (\lambda_i)$ and in 
$\delta_{\rm EM}^{K \ell}$  well  below the percent 
level~\cite{flavianet}. We  can safely ignore this source of uncertainty 
in $\delta_{\rm EM}^{K \ell}$. 

\item  To the order we work, the electromagnetic LECs contribute a $v$-independent  term 
to $\delta f_{\pm}^{K \pi}(v)$, which thus affects the decay rates as follows,
\beq
\delta_{\rm EM}^{K \ell} ({\cal D}_3) \ \sim \ 
2 \, \delta f_+ \Big|_{\rm LECs}    \cdot 
\frac{\displaystyle \int_{{\cal D}_3}\, dy dz  A_1^{(0)} (y,z)  }{I_{K \ell}^{(0)} (\lambda_i)} 
\ + \  \delta f_- \Big|_{\rm LECs}    \cdot 
\frac{ \displaystyle \int_{{\cal D}_3}\, dy dz  A_2^{(0)} (y,z)  }{I_{K \ell}^{(0)} (\lambda_i)} ~. 
\eeq  
The coefficient of  $2 \, \delta f_+$ is $O(1)$ while the coefficient of 
$\delta f_{-}$ is  roughly $0.2$ for $K_{\mu 3}$ decays and completely negligible 
for $K_{e 3}$ decays, being $O(m_e/M_K)^2$. 
Taking a very conservative attitude, we assign  a $100 \%$ fractional uncertainty  
 to $\delta f_{\pm} |_{\rm LECs}$  (using LECs central values from  
 Refs.~\cite{DescotesGenon:2005pw,mouss-ana}), 
 which induces an absolute uncertainty in all the $\delta_{\rm EM}^{K \ell} $  of about 
  $\pm  0.1 \%$. 

\end{itemize}

In order to discuss the  error coming from 
higher order chiral corrections not included in our analysis,  
we find it useful to decompose    (to each chiral order) 
the EM corrections $\delta^{K \ell}$ in terms of $\delta_{1,2,3, 4}$:
\begin{eqnarray}
\delta^{K^0 e}  &=&   \delta_1 + \delta_2 + \delta_3 + \delta_4 ~, \nonumber  \\
\delta^{K^0 \mu}   &=&   \delta_1 + \delta_2 - \delta_3 - \delta_4 ~, \nonumber  \\
\delta^{K^\pm e}   &=&   \delta_1 - \delta_2 +\delta_3 - \delta_4 ~, \nonumber  \\
\delta^{K^\pm  \mu}  &=&   \delta_1 - \delta_2 - \delta_3 + \delta_4 ~. 
\end{eqnarray}
Here  $\delta_{1}$ represents  a correction common to all modes, 
$\delta_2$  a correction anti-correlated in kaon isospin but blind to lepton flavor,  
and finally $\delta_{3,4}$  are lepton-universality breaking terms, correlated and 
anti-correlated in kaon isospin,  respectively. 
To  $O(e^2 p^0)$ we find  $\delta_1^{e^2 p^0} = 0.63 \%$,   $\delta_2^{e^2 p^0} = 0.57 \%$,  
$\delta_3^{e^2 p^0} = -0.08  \%$,  $\delta_4^{e^2 p^0} =  - 0.12 \%$.  
On the basis of chiral power counting  we expect  the higher order corrections to scale as 
\begin{equation}
 \delta_i^{e^2 p^2}  \sim    (M_K/(4 \pi  F_\pi))^2 \cdot   \delta_i^{e^2 p^0}  
\sim  0.2  \cdot   \delta_i^{e^2 p^0}~~.           
\end{equation}
This estimate is validated by comparison of the fixed chiral order results 
with the ones obtained within  the \lq\lq soft-photon factorization" approach 
discussed in Section 4 of Ref.~\cite{Cirigliano:2001mk}, 
which  include a class of  higher order chiral corrections (see
Table~\ref{tab:2001}) ~\footnote{ 
Note that in  Section 5.3  of  Ref.~\cite{Cirigliano:2001mk} an alternative factorization prescription 
is given, which is  valid only for $K_{e3}$ modes. 
The latter  prescription was used in the numerical analysis of 
Refs. ~\cite{Cirigliano:2001mk,Cirigliano:2004pv} and would lead  to results slightly different 
from  those of Table~\ref{tab:2001}.  The first two entries in the first column should be 
replaced as follows:  $0.41 \to 0.56 $ and $ -0.564 \to -0.41$. }.
The only anomaly appears to be in the coefficient $\delta_3$, where 
we find $\delta_3:  \ - 0.08 \% \to -0.16 \%$  when going from fixed chiral order to 
the soft factorization scheme. 
This can be traced back to the  cancellation between the negative contribution 
from ${\cal D} _3$  (-0.31$\%$) and the positive contribution from
${\cal D}_{4-3}$ (0.23$\%$). Multiplying these individual pieces by 0.2 gives
$\sim  0.06$ and $\sim 0.05$, respectively, which is just the order of magnitude of the 
shift we are seeing (-0.08 $\to $ -0.16). 

Based on the above discussion,  we bound the higher order uncertainties as follows: 
 $|\delta_1^{e^2 p^2}| <  0.13 \%$,  
$|\delta_2^{e^2 p^2}| <  0.11 \%$,  
$|\delta_3^{e^2 p^2}| <  0.08 \%$,  
$|\delta_4^{e^2 p^2}| <  0.025 \%$. 
Adding these linearly we estimate the uncertainties quoted in 
Table~\ref{tab:chiral} for the total corrections.     
Finally,  using the same bounds on $\delta_i^{e^2 p^2}$  we estimate 
the theoretical uncertainties on the linear combinations which 
are relevant for lepton universality and strong isospin-breaking  tests: 
\bea
\delta_{EM}^{K^0 e} -  \delta_{EM}^{K^0 \mu} & =&   - (0.41 \pm 0.20) \%  ~,  \\
\delta_{EM}^{K^\pm e} -  \delta_{EM}^{K^\pm  \mu} & =&    (0.08 \pm 0.20) \% ~,   \\
\delta_{EM}^{K^\pm e} -   \delta_{EM}^{K^0 e}  &=& - (0.89 \pm 0.30) \% ~, \\ 
\delta_{EM}^{K^\pm \mu} -   \delta_{EM}^{K^0 \mu}  &=& - (1.38 \pm 0.30) \% ~.
\eea 
%

%%%%%%%%%%%%%%%%%%%%% Table    %%%%%%%%%%%%%%%%% 
\begin{table}[t!]
\begin{center}
\begin{tabular}{|c|c|c|c|c|}
\hline
  &  $I^{(0)}_{K\ell} (\lambda_i)$  &   $\delta_{\rm EM}^{K \ell} ({\cal D}_3) (\%)$  &
   $\delta_{\rm EM}^{K \ell} ({\cal D}_{4-3}) (\%) $   &    $\delta_{\rm EM}^{K \ell}  (\%)$  \\[5pt]
\hline
$K^0_{e3} $ & 0.103070  & 0.50  &  0.49 &  0.99 $\pm $ 0.30 \\
$K^\pm_{e3}$  & 0.105972   & -0.35 &  0.45  & 0.10 $\pm $ 0.30  \\
$K^0_{\mu 3}$  & 0.068467  & 1.38  &   0.02  & 1.40 $\pm $ 0.30  \\
$K^\pm_{\mu 3}$  & 0.070324  & 0.007 & 0.009 & 0.016 $\pm $ 0.30  \\
\hline
\end{tabular}
\end{center}
\caption{Summary of phase space integrals 
and EM corrections to the $K_{\ell 3}$ 
decay  rates.  The EM corrections are calculated to fixed order in ChPT ($O(e^2 p^0)$).
The phase space integrals are calculated using 
slope and curvature parameters from the fit of Ref.~\protect\cite{flavianet}.  
The uncertainty estimate is discussed in the text. } 
\label{tab:chiral}
\end{table}
%%%%%%%%%%%%%%%%%%%%%%%%%%%%%%%%%%%%%%%%%%%

%%%%%%%%%%%%%%%%%%%%% Table    %%%%%%%%%%%%%%%%% 
\begin{table}[t!]
\begin{center}
\begin{tabular}{|c|c|c|c|}
\hline
% &  $I^{(0)}_{K\ell}$ 
 &   $\delta_{\rm EM}^{K \ell} ({\cal D}_3) (\%)$  &
   $\delta_{\rm EM}^{K \ell} ({\cal D}_{4-3}) (\%) $   &    $\delta_{\rm EM}^{K \ell}  (\%)$  \\[5pt]
\hline
$K^0_{e3} $ 
& 0.41  &  0.59 &  1.0  \\
$K^\pm_{e3}$  
& -0.564 &  0.528  & -0.04  \\
$K^0_{\mu 3}$  
 & 1.57  &   0.04  & 1.61  \\
$K^\pm_{\mu 3}$  
& -0.006 & 0.011 & 0.005  \\
\hline
\end{tabular}
\end{center}
\caption{Summary of  EM corrections to the $K_{\ell 3}$ 
decay  rates  calculated according to the 
\lq\lq soft-photon factorization" approach of Ref.~\protect\cite{Cirigliano:2001mk}, which includes 
{\em incomplete} higher order terms in the chiral expansion.   
Comparison with the results of Table~\ref{tab:chiral} validates our estimate 
of the theoretical uncertainties.} 
\label{tab:2001}
\end{table}
%%%%%%%%%%%%%%%%%%%%%%%%%%%%%%%%%%%%%%%%%%%

\section{Conclusions}
\label{sect:conclusions}
In this work we have provided a unified discussion of the 
radiative corrections of $O(e^2 p^0)$ 
to  all $K_{\ell 3}$ decay rates.   
We have argued that through the  calculation of $K \to \pi \ell \nu$ 
amplitudes to $O(e^2p^2)$ in the chiral effective theory  
we can derive the complete corrections of 
$O(e^2 p^0)$ to  the Dalitz plot density and  the integrated decay rate.
We have systematically discarded higher order effects that are only partially known, 
and we have included the unknown effects in a generous estimate of the 
theoretical uncertainty.     
For the first time we have presented  complete numerical results for the $K_{\mu 3}$ modes, 
while also updating the previous analysis of $K_{e3}$ modes. 

The main outcome of our investigation is summarized in Table~\ref{tab:chiral}, 
which contains the corrections  to the total (fully photon inclusive) decay rates of all  
$K_{\ell 3}$ decay modes ($K=K^\pm,K^0; \ell= e,\mu $). 
These results provide important theoretical input for the determination of 
the product  $f_{+}^{K^0 \pi^-} (0) \cdot V_{us}$  
from $K_{\ell 3}$ 
decays  at the $0.2 \%$ level~\cite{flavianet} through Eq.~\ref{eq:master}.  
This task requires  as additional theoretical input the factor $\delta_{SU(2)}^{K \pi}$, 
which has been recently updated   in Refs.~\cite{bijnens07,helmut}. 
We refrain here from producing a number for $f_{+}^{K^0 \pi^-} (0) \cdot V_{us}$:
the result reflecting  most recent experimental data 
can be found in the Flavianet Kaon Working Group web page~\cite{flavianet-web}. 

Further reduction of the theoretical uncertainty on the corrections 
$\delta^{\rm EM}_{K \ell}$ would require an analysis of the amplitudes 
 to $O(e^2 p^4)$ in ChPT, which is beyond the scope of this work. 
 At the moment there  appears to be no immediate need for such an analysis, since the 
error  on  $V_{us}$ is  dominated by the $\sim 1 \%$ theoretical uncertainty in  
 $f_{+}^{K^0 \pi^-} (0)$,   for which a compilation and discussion of theoretical results can  be 
 found in Ref.~\cite{flavianet}. \\

{\bf  Acknowledgements} --   The work of V.C.  is  supported by the 
U.S.  DOE  Office of Science  and by the LDRD 
Program at Los Alamos National Laboratory. 
H.N. thanks Julia Schweizer for collaboration at an early stage of this 
work.  V.C. thanks Gino Isidori for useful discssions. 
We thank Alex Friedland for helping us handle large figure files.

% \newpage

\appendix

\section{Corrections induced by virtual photons: $\delta f_{\pm}^{K \pi} (v)$} 
\label{app:loops}

Using the notation of Refs.~\cite{Cirigliano:2001mk,Cirigliano:2004pv}, 
and identifying $f_{\pm}^{\rm EM-loc}$ of Ref.~\cite{Cirigliano:2001mk} 
with $\widehat{f}_{\pm}$ of  Ref.~\cite{Cirigliano:2004pv}, 
we have:
\bea
\delta f_{+}^{K^+ \pi^0} (u)  &=&  
\frac{\alpha}{4 \pi}  \Bigg[
\Gamma_1 (u, m_\ell^2, M_K^2) +  \Gamma_2  (u, m_\ell^2, M_K^2) 
\Bigg] 
+  \widehat{f}_{+}^{\rm K^+ \pi^0} ~,
\ \ \ \ \ 
\\
\delta f_{-}^{K^+ \pi^0} (u)  &=&  
\frac{\alpha}{4 \pi}  \Bigg[
\Gamma_1 (u, m_\ell^2, M_K^2) -  \Gamma_2  (u, m_\ell^2, M_K^2) 
\Bigg]
+ \widehat{f}_{-}^{\rm K^+ \pi^0}   ~, 
\eea
and
\bea
\delta f_{+}^{K^0 \pi^-} (s)  &=&  
\frac{\alpha}{4 \pi}  \Bigg[
\Gamma_1 (s, m_\ell^2, M_\pi^2) +  \Gamma_2  (s, m_\ell^2, M_\pi^2) 
\Bigg]
+  \widehat{f}_{+}^{\rm K^0 \pi^- } ~,   
\ \ \ \ \ 
\\
\delta f_{-}^{K^0 \pi^-} (s)  &=&  
\frac{\alpha}{4 \pi}  \Bigg[
\Gamma_2 (s, m_\ell^2, M_\pi^2) -  \Gamma_1  (s, m_\ell^2, M_\pi^2) 
\Bigg]
+ \widehat{f}_{-}^{\rm K^0 \pi^- }  ~.
\eea

%  \newpage

%%%%%%%%%%%%%%%%%

%%%%%%%%%%%%%%%%%


\begin{thebibliography}{99}



\bibitem{flavianet}
  M.~Antonelli {\it et al.}  [FlaviaNet Working Group on Kaon Decays],
  %``Precision tests of the Standard Model with leptonic and semileptonic kaon
  %decays,''
  arXiv:0801.1817 [hep-ph].
  %%CITATION = ARXIV:0801.1817;%%


\bibitem{chpt}
S. Weinberg, Physica A {\bf 96}, 327 (1979);
J. Gasser and H. Leutwyler, Ann. Phys. {\bf 158}, 142 (1984);
H. Leutwyler, Ann. Phys. {\bf 235}, 165 (1994) 

\bibitem{photons}
R. Urech, Nucl. Phys. B {\bf 433}, 234 (1995);
H. Neufeld and H. Rupertsberger, Z. Phys. C {\bf 68}, 91 (1995);
H. Neufeld and H. Rupertsberger, Z. Phys. C {\bf 71}, 131 (1996).


%\cite{Knecht:1999ag}
\bibitem{Knecht:1999ag}
  M.~Knecht, H.~Neufeld, H.~Rupertsberger and P.~Talavera,
  %``Chiral perturbation theory with virtual photons and leptons,''
  Eur.\ Phys.\ J.\ C {\bf 12}, 469 (2000)
  [arXiv:hep-ph/9909284].


%\cite{Cirigliano:2001mk  \cite{Cirigliano:2004pv}}
\bibitem{Cirigliano:2001mk}
  V.~Cirigliano, M.~Knecht, H.~Neufeld, H.~Rupertsberger and P.~Talavera,
  %``Radiative corrections to K(l3) decays,''
  Eur.\ Phys.\ J.\  C {\bf 23}, 121 (2002)
  [arXiv:hep-ph/0110153].
  %%CITATION = EPHJA,C23,121;%%


%\cite{Cirigliano:2004pv}
\bibitem{Cirigliano:2004pv}
  V.~Cirigliano, H.~Neufeld and H.~Pichl,
  %``K(e3) decays and CKM unitarity,''
  Eur.\ Phys.\ J.\  C {\bf 35}, 53 (2004)
  [arXiv:hep-ph/0401173].
  %%CITATION = EPHJA,C35,53;%%



%\cite{DescotesGenon:2005pw}
\bibitem{DescotesGenon:2005pw}
  S.~Descotes-Genon and B.~Moussallam,
  %``Radiative corrections in weak semi-leptonic processes at low energy: A
  %two-step matching determination,''
  Eur.\ Phys.\ J.\  C {\bf 42}, 403 (2005)
  [arXiv:hep-ph/0505077].
  %%CITATION = EPHJA,C42,403;%%

\bibitem{mouss-ana}  
%%\cite{Ananthanarayan:2004qk}
%\bibitem{Ananthanarayan:2004qk}
  B.~Ananthanarayan and B.~Moussallam,
  %``Four-point correlator constraints on electromagnetic chiral parameters  and
  %resonance effective Lagrangians,''
  JHEP {\bf 0406}, 047 (2004)
  [arXiv:hep-ph/0405206].
  %%CITATION = JHEPA,0406,047;%%
  
\bibitem{HN-frascati07} 
H.~Neufeld,  talk presented at 
\lq\lq  Mini-Workshop on K physics",  18-19 May 2007, INFN Frascati, Italy  
(available at http://www.lnf.infn.it).

\bibitem{VC-kaon07} 
V.~ Cirigliano,  PoS (KAON): 007, 2007 (available at http://pos.sissa.it).

\bibitem{Post:2001si} % \bibitem{Bijnens:2003uy}
P.~Post and K.~Schilcher,
%``K(l3) form factors at order p**6 in chiral perturbation theory,''
Eur.\ Phys.\ J.\ C {\bf 25} (2002) 427
[hep-ph/0112352].
%%CITATION = HEP-PH 0112352;%%

\bibitem{Bijnens:2003uy}
J.~Bijnens and P.~Talavera,
%``K(l3) decays in chiral perturbation theory,''
Nucl.\ Phys.\ B {\bf 669} (2003) 341
[hep-ph/0303103]; see also 
http://www.thep.lu.se/$\sim$bijnens/chpt.html.
%%CITATION = HEP-PH 0303103;%%

\bibitem{bijnens07}
%\cite{Bijnens:2007xa}
%\bibitem{Bijnens:2007xa}
 J.~Bijnens and K.~Ghorbani,
%``Isospin breaking in $K\pi$ vector form-factors for the weak and rare decays
%$K_{\ell3}$, $K\to\pi\nu\bar\nu$ and $K\to\pi\ell^+\ell^-$,''
arXiv:0711.0148 [hep-ph].
%%CITATION = ARXIV:0711.0148;%%



\bibitem{f0-analytic}
%\cite{Leutwyler:1984je}
%\bibitem{Leutwyler:1984je}
  H.~Leutwyler and M.~Roos,
  %``Determination Of The Elements V(Us) And V(Ud) Of The Kobayashi-Maskawa
  %Matrix,''
  Z.\ Phys.\  C {\bf 25}, 91 (1984);  
  %%CITATION = ZEPYA,C25,91;%%
%
%\cite{Jamin:2004re}
%\bibitem{Jamin:2004re}
  M.~Jamin, J.~A.~Oller and A.~Pich,
  %``Order p**6 chiral couplings from the scalar K pi form factor,''
  JHEP {\bf 0402}, 047 (2004)
  [arXiv:hep-ph/0401080]; 
  %%CITATION = JHEPA,0402,047;%%
%
%\cite{Cirigliano:2005xn}
%\bibitem{Cirigliano:2005xn}
  V.~Cirigliano, G.~Ecker, M.~Eidem\"uller, R.~Kaiser, A.~Pich and 
J.~Portoles,
  %``The  Green function and SU(3) breaking in K(l3) decays,''
  JHEP {\bf 0504}, 006 (2005)
  [arXiv:hep-ph/0503108].
  %%CITATION = JHEPA,0504,006;%%


\bibitem{f0-lattice}   
D.~Be\'cirevi\'c {\it et al.}, Nucl. Phys. B {\bf 705}, 339  (2005) [hep-ph/0403217] ; 
%%CITATION = HEP-PH 0403217;%%
  M.~Okamoto  [Fermilab Lattice Collaboration],
  %``Full CKM matrix with lattice QCD,''
  hep-lat/0412044 ; 
  %%CITATION = HEP-LAT 0412044;%%
  N.~Tsutsui {\it et al.}  [JLQCD Collaboration],
  %``Kaon semileptonic decay form factors in two-flavor QCD,''
  Proc.\ Sci.\  {\bf LAT2005} (2005) 357
  [hep-lat/0510068] ;
  %%CITATION = HEP-LAT 0510068;%%
  C.~Dawson, T.~Izubuchi, T.~Kaneko, S.~Sasaki and A.~Soni,
%   ``Vector form factor in K(l3) semileptonic decay with two flavors of
  %dynamical domain-wall quarks,''
  arXiv:hep-ph/0607162; 
  %%CITATION = HEP-PH 0607162;%%
%
  D.~J.~Antonio {\it et al.},
  %``K --> pi l nu form factor with N(f) = 2+1 dynamical domain wall fermions,''
  arXiv:hep-lat/0610080; 
  %%CITATION = HEP-LAT 0610080;%%
%\cite{Boyle:2007qe}
%\bibitem{Boyle:2007qe}
  P.~A.~Boyle {\it et al.},
  %``Kl3 semileptonic form factor from 2+1 flavour lattice QCD,''
  Phys.\ Rev.\ Lett.\  {\bf 100}, 141601 (2008)
  [arXiv:0710.5136 [hep-lat]].
  %%CITATION = PRLTA,100,141601;%%

 \bibitem{sternetal}
 %\cite{Bernard:2006gy}
%\bibitem{Bernard:2006gy}
  V.~Bernard, M.~Oertel, E.~Passemar and J.~Stern,
  %``K(L)(mu3) decay: A stringent test of right-handed quark currents,''
  Phys.\ Lett.\  B {\bf 638}, 480 (2006)
  [arXiv:hep-ph/0603202].
  %%CITATION = PHLTA,B638,480;%%


%\cite{Sirlin:1977sv}
\bibitem{Sirlin-sew}
  A.~Sirlin,
  %``Current Algebra Formulation Of Radiative Corrections In Gauge Theories And
  %The Universality Of The Weak Interactions,''
  Rev.\ Mod.\ Phys.\  {\bf 50}, 573 (1978)
  [Erratum-ibid.\  {\bf 50}, 905 (1978)]; 
  %%CITATION = RMPHA,50,573;%%
%\cite{Sirlin:1981ie}
%\bibitem{Sirlin:1981ie}
  A.~Sirlin,
  %``Large M(W), M(Z) Behavior Of The O(Alpha) Corrections To Semileptonic
  %Processes Mediated By W,''
  Nucl.\ Phys.\  B {\bf 196}, 83 (1982); 
  %%CITATION = NUPHA,B196,83;%%
%\cite{Marciano:1993sh}
%\bibitem{Marciano:1993sh}
  W.~J.~Marciano and A.~Sirlin,
  %``Radiative corrections to pi(lepton 2) decays,''
  Phys.\ Rev.\ Lett.\  {\bf 71}, 3629 (1993).
  %%CITATION = PRLTA,71,3629;%%

\bibitem{IR-YFS-W}  
%\cite{Yennie:1961ad}
%\bibitem{Yennie:1961ad}
  D.~R.~Yennie, S.~C.~Frautschi and H.~Suura,
  %``The infrared divergence phenomena and high-energy processes,''
  Annals Phys.\  {\bf 13} (1961) 379;
  %%CITATION = APNYA,13,379;%%
%\cite{Weinberg:1965nx}
%\bibitem{Weinberg:1965nx}
  S.~Weinberg,
  %``Infrared photons and gravitons,''
  Phys.\ Rev.\  {\bf 140} (1965) B516.
  %%CITATION = PHRVA,140,B516;%%

\bibitem{ginsberg} 
%\cite{Ginsberg:1968pz}
% \bibitem{Ginsberg:1968pz}
  E.~S.~Ginsberg,
  %``Radiative corrections to k-e-3-neutral decays and the delta-i=1/2 rule.
  %(erratum),''
  Phys.\ Rev.\  {\bf 171} (1968) 1675
  [Erratum-ibid.\  {\bf 174} (1968) 2169]; 
  %%CITATION = PHRVA,171,1675;%%
%\cite{Ginsberg:1969jh}
%\bibitem{Ginsberg:1969jh}
  E.~S.~Ginsberg,
  %``Radiative corrections to the k-l-3 +- dalitz plot. (erratum),''
  Phys.\ Rev.\  {\bf 162} (1967) 1570
  [Erratum-ibid.\  {\bf 187} (1969) 2280];  
  %%CITATION = PHRVA,162,1570;%%
%\cite{Ginsberg:1970vy}
% \bibitem{Ginsberg:1970vy}
  E.~S.~Ginsberg,
  %``Radiative corrections to k-mu-3 decays,''
  Phys.\ Rev.\  D {\bf 1} (1970) 229.
  %%CITATION = PHRVA,D1,229;%%

\bibitem{rambos}
%\cite{Kleiss:1985gy}
%\bibitem{Kleiss:1985gy}
  R.~Kleiss, W.~J.~Stirling and S.~D.~Ellis,
  %``A New Monte Carlo Treatment Of Multiparticle Phase Space At
  %High-Energies,''
  Comput.\ Phys.\ Commun.\  {\bf 40} (1986) 359.
  %%CITATION = CPHCB,40,359;%%

\bibitem{pdg}
%\cite{Yao:2006px}
%\bibitem{Yao:2006px}
  W.~M.~Yao {\it et al.}  [Particle Data Group],
  %``Review of particle physics,''
  J.\ Phys.\ G {\bf 33} (2006) 1.
  %%CITATION = JPHGB,G33,1;%%

\bibitem{jackson} J.~D.~ Jackson,  {\it Classical Electrodynamics}, 3rd 
ed., New York: Wiley.


\bibitem{helmut}
%\cite{Kastner:2008ch}
%\bibitem{Kastner:2008ch}
  A.~Kastner and H.~Neufeld,
  %``The Kl3 scalar form factors in the standard model,''
  arXiv:0805.2222 [hep-ph].
  %%CITATION = ARXIV:0805.2222;%%

\bibitem{flavianet-web} The Flavianet Kaon Working Group maintains updated information 
at the following URL:  http://www.lnf.infn.it/wg/vus. 



\end{thebibliography}
\end{document}